\documentclass[preprint,showpacs,preprintnumbers]{revtex4}%

\usepackage{amsfonts}
\usepackage{amsmath}
\usepackage{amssymb}
\usepackage{graphicx}%
\begin{document}
\title{Sign reversal of Hall conductivity and quantum confinement in graphene ribbons}
\author{Junfeng Liu and Zhongshui Ma}
\affiliation{School of Physics, Peking University, Beijing 100871, China}

\begin{abstract}
Characterized by zigzag and armchair boundaries, the narrow
ribbons display the very different characteristics in Hall
conductivities. It is shown that the multi-band-crossings occur in
the energy spectrum for armchair ribbons, and the number of them
depends on the width of ribbons. Theoretically, it is predicated
that the conductivities exhibit drastic sign reversals for narrow
ribbons as the Fermi energy sweep over the band-crossings. A new
classification of armchair ribbons is suggested based on the
emergence of a flat band in the energy spectrum only for odd
armchair ribbons. The evolution of jumped Hall conductivities to
step-like plateaus and the restore of density of states at the van
Hove singularity in the limitation to graphene sheet have been
analyzed.

\end{abstract}
\maketitle

Experimentally, many novel
properties\cite{novo1,novo2,zhang1,berg} have been observed in
graphene systems, such as the electron-hole symmetry, the
odd-integer quantum Hall effect\cite{novo2,zhang1}, and the finite
conductivity in dissipationless process\cite{novo2}. Among these
the peculiar odd-integer Hall plateaus has been theoretically
recognized in the frame of the Landau quantization of massless
Dirac fermions\cite{jw,yz,vp,ks,nmr}. Considering the potential
applications in future nanoelectronic devices, there is a steep
rise recently in the interest in studying the various properties
of graphene nanoribbons (GNRs). Borne on their honeycomb atomic
structure, two kinds of ribbons, with zigzag edges (ZGNRs) and
armchair edges (AGNRs) respectively, can be configured when
graphene sheets are transversely cut. Actually, transversely
reduced size of ribbons introduces important physical phenomena
such as quantum confinement and edge effects, which have been
revealed in the studying of electronic structure, unconventional
transport, and optical excitations\cite{obra,ouy,yan}. The
anisotropy in their electrical properties relies upon the
structures of GNRs in the special set up\cite{lg,lei,mmu,cheo}.
Therefore, of particular interest, it is desirable to know how the
effects of the quantum confinement and the edge characteristics
are reflected in the Hall-like conductivity in GNRs under a
magnetic field.

In the present letter, we consider narrow ribbons subject to a
perpendicular magnetic field for this purpose. We take some values
of the strength of magnetic field so that Landau levels are not
well developed in the comparison of the characteristic
discretization of energy spectrum originated from the transversely
confined sizes of ribbons. The magnetic field is, here, used to
play only a role to achieve the electronic motion in longitudinal
direction of GNRs. Hall-like conductivities can, then, be
investigated. Because multi-band-crossings occur in the energy
spectrum for AGNRs while only one at van Hove singularity for
ZGNRs, the behaviors of electrical properties under the magnetic
field are expected to be very different for ZGNRs and AGNRs when
Fermi energy sweeps over the band-crossings. We show that, similar
to the infinite graphene sheet, the Hall conductivities for ZGNRs
undergo a sign-reversal jump merely at the van Hove singularities
where the transition from the Dirac fermion to the ordinary
fermion happens\cite{yh1,yh2}. But, the Hall conductivities for
AGNRs exhibit extra drastic sign-reversal jumps as functions of
Fermi energy. The number of sign reversals depends on the width of
AGNRs. The regularity is not changed regardless of AGNRs being
metal or semiconductor. This reflects the physical situation in
which more than one meaningful classifications are involved to
characterize the electronic structure for ribbons. These extra
sign-reversal jumps in the Hall conductivity for AGNRs could be
receded when the width of ribbons or the strength of magnetic
field is increased. In the limitation of infinite graphene sheets,
the oscillations would disappear except that one related to the
van Hove singularity and the step-like plateaus are formed, which
is consistent with observations for the infinite graphene
sheet\cite{yh1,yh2}.

We describe GNRs by a tight-binding Hamiltonian on two-dimensional (2D)
honeycomb lattice $H=t\sum_{\left\langle ij\right\rangle }\exp\left(
i\gamma_{ij}\right)  c_{i}^{\dagger}c_{j}$, where $\left\langle
i,j\right\rangle $ denotes the summation over the nearest neighbor sites, $t$
($=2.71eV$) is the hopping integral for nearest neighbors, and $c_{i}%
^{\dagger}$ ($c_{i}$) represents the creation (annihilation) operator of
electrons on the site $i$ neglecting the spin degree of freedom. The magnetic
field is applied perpendicularly to the sheet of the ribbon and is responsible
for incorporating the Peierls phase $\gamma_{ij}=\left(  2\pi/\phi_{0}\right)
\int_{i}^{j}\mathbf{A}\cdot d\mathbf{l}$, with the vector potential
$\mathbf{A}$ and the magnetic flux quantum $\phi_{0}=hc/e$. ZGNRs and AGNRs
are classified by their characteristics on edges, respectively\cite{fu3}. The
width $N$ of ZGNRs is defined by the number of longitudinal zigzag lines,
while by dimer lines for AGNRs. In our calculations, we extend the coupled
Harper equations to describe a ribbon in width $N$ with the boundary condition
$\psi_{0}=\psi_{N+1}=0$. The corresponding Hamiltonian can, then, be written
as a $2N\times2N$ matrix\cite{liu}. For a ribbon along the $y$-direction
longitudinally and its transversal section along the $x$-direction, the
eigenfunction can be expressed in the form of $\psi_{kj}=\xi_{j}(k)\exp\left(
iky\right)  $, where $k=k_{y}$ is the longitudinal wave vector while $j$
($j=1,2,3,\cdots2N$) indicate the transversal channels on the transversal
section, and $\xi_{kj}$ denotes the $j$-th eigenstate which satisfies equation
$H\xi_{kj}=\epsilon_{kj}\xi_{kj}$. The energy eigenvalues $\epsilon_{kj}$ and
the eigenvectors $\xi_{kj}$ can be obtained by a numerical diagonalization of
the Hamiltonian. As a finite system, the energy levels are discretized.

To calculate the DC Hall conductivity at zero temperature we apply Kubo
formula \cite{liu2}, $\sigma_{yx}=-\left(  2\hbar/W\right)  \sum_{k}%
\sum_{\epsilon_{kj}>E_{F},\epsilon_{kj^{\prime}}<E_{F}}Im\left(
J_{jj^{\prime}}^{y}J_{j^{\prime}j}^{x}\right)  /\left(  \epsilon_{kj}%
-\epsilon_{kj^{\prime}}\right)  ^{2}$, where $W$ is the width of
ribbons, $W=(3N/2-1)a$ for ZGNRs and $W=\left(  \sqrt{3}a/2\right)
(N-1)$ for AGNRs with the c-c bond length $a=0.142nm$. The current
operators $J_{\alpha}$ ($\alpha=x$ and $y$) are obtained by
$J_{\alpha}=c\left(  \partial H/\partial A_{\alpha}\right)  $. In
the Hilbert space of eigenvectors $\xi_{j}$, current operators can
be represented in terms of a matrix with elements $J_{jj^{\prime
}}^{\alpha}=\xi_{j}^{\dag}J_{\alpha}\xi_{j^{\prime}}$, where
$\xi_{j}$ and $\epsilon_{kj}$ are the $j$-th eigenstate and
eigenvalue. The summation in this formula denotes that $j$ always
takes in the subbands above the Fermi energy while $j^{\prime}$
always in the subbands below the Fermi energy. Thus, if these two
subbands $j$ and $j^{\prime}$ are crossed, the indices $j $ and
$j^{\prime}$ would undergo an exchange when the Fermi energy
sweeps over the energy at the crossing point. The contributions to
$\sigma_{yx}$ from the two crossing subbands becomes divergence if
the Fermi energy comes close to the crossing point. Therefore, the
exchange between crossing bands $j$ and $j^{\prime}$ leads to a
sign reversal and the drastic jump in the Hall-like conductivity.

\begin{figure}[ptb]
\begin{center}
\includegraphics[bb=15 11 291 239, width=3.005in]{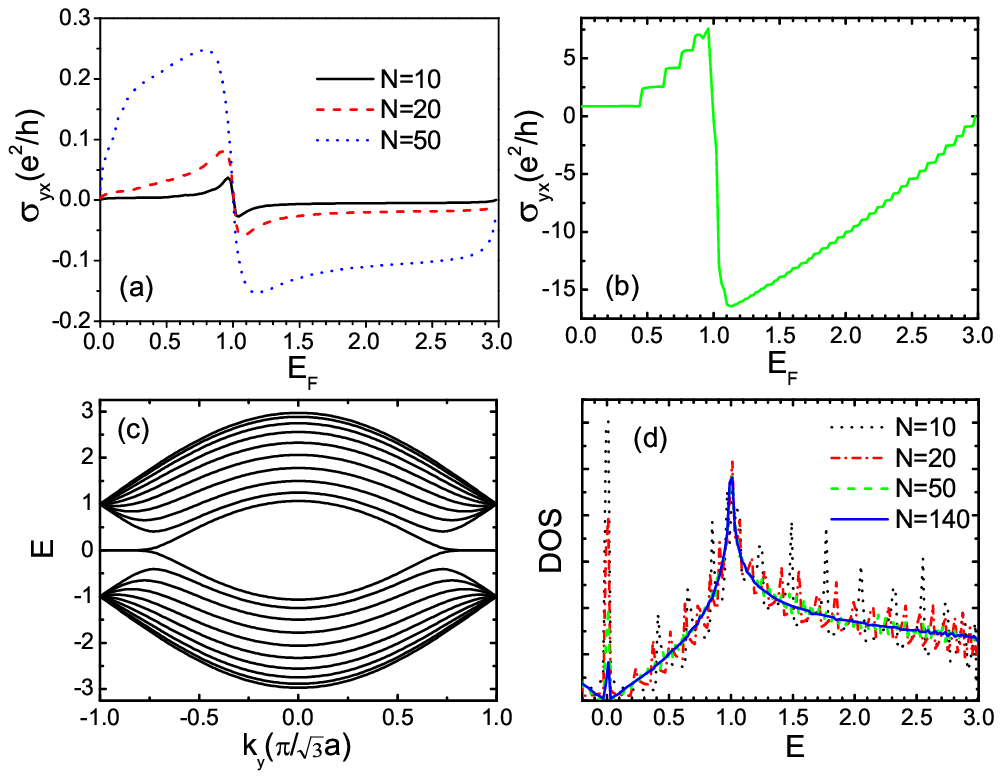}
\end{center}
\caption{(a) The Hall conductivity versus the Fermi energy for
ZGNRs at the zero temperature: widths $N=10$, $20$, and $50$ are
chosen. The magnetic flux through a plaquette in the unit of a
quantum flux, $f=1.0\times10^{-4}$; (b) The Hall conductivity in
the strong field regime, where $N=60$ and $f=2.0\times10^{-2}$;
(c) The energy spectrum for a ZGNR in width $N=10$; and
(d) The evolution of DOS with widening ribbons.}%
\label{Liufig1}%
\end{figure}

Now we investigate the characteristics in Hall-like conductivities
for ZGNRs and AGNRs, respectively. In our numerical calculations
all energies are taken in the unit of hopping integral and the
magnetic field is introduced by the magnetic flux through a
plaquette in the unit of a quantum flux, $f\equiv \phi /\phi
_{0}$. The strength of magnetic field is chosen not to destroy the
band-crossings caused by transverse confinement. Therefore, the
band-structure is dominated by its transverse confinement.
Electrons move along longitudinal direction when a transverse
electric field is applied. For the ZGNRs, the numerical
calculation shows that band-crossings appear only at
$E=\pm\left\vert t\right\vert $\cite{fu1,fu2}
(Fig.\ref{Liufig1}(c)). Correspondingly, the Hall conductivity,
$\sigma_{yx}$, exhibits the
sign-reversal jumps at $E=\pm\left\vert t\right\vert $. In Fig.\ref{Liufig1}%
(a) we have shown the Hall conductivities for ZGNRs in various widthes under
the magnetic flux $f=1.0\times10^{-4}$ via the Fermi energy over the region
$0<E_{F}<3\left\vert t\right\vert $. For $-3\left\vert t\right\vert <E_{F}<0$
the behavior of $\sigma_{yx}$ is antisymmetric to that for $0<E_{F}%
<3\left\vert t\right\vert $ due to the electron-hole symmetry.
Actually, the appearance of jumps at $E=\pm\left\vert t\right\vert
$ is robust and attributed to the van Hove singularity as happened
for infinite graphene sheets\cite{yh1,yh2}. Increasing the
strength of magnetic field or widening the ribbons lead the Hall
conductivity gradually to become quantized, as shown in
Fig.\ref{Liufig1}(b). The reason is that if the effect of magnetic
field is strong enough, the Landau levels dominate the band
structure. It, thus, leads to the quantized Hall conductivity,
which is equivalent to increasing the width of ribbon at a fixed
magnetic field. When width of ribbon is increased large enough,
the discretization of bands is receded and Landau levels become
dominative. The quantized plateaus in the Hall conductivity are
achieved as experimental observation for infinite graphene sheets.

\begin{figure}[ptb]
\begin{center}
\includegraphics[bb=15 17 286 386, width=3.205in]{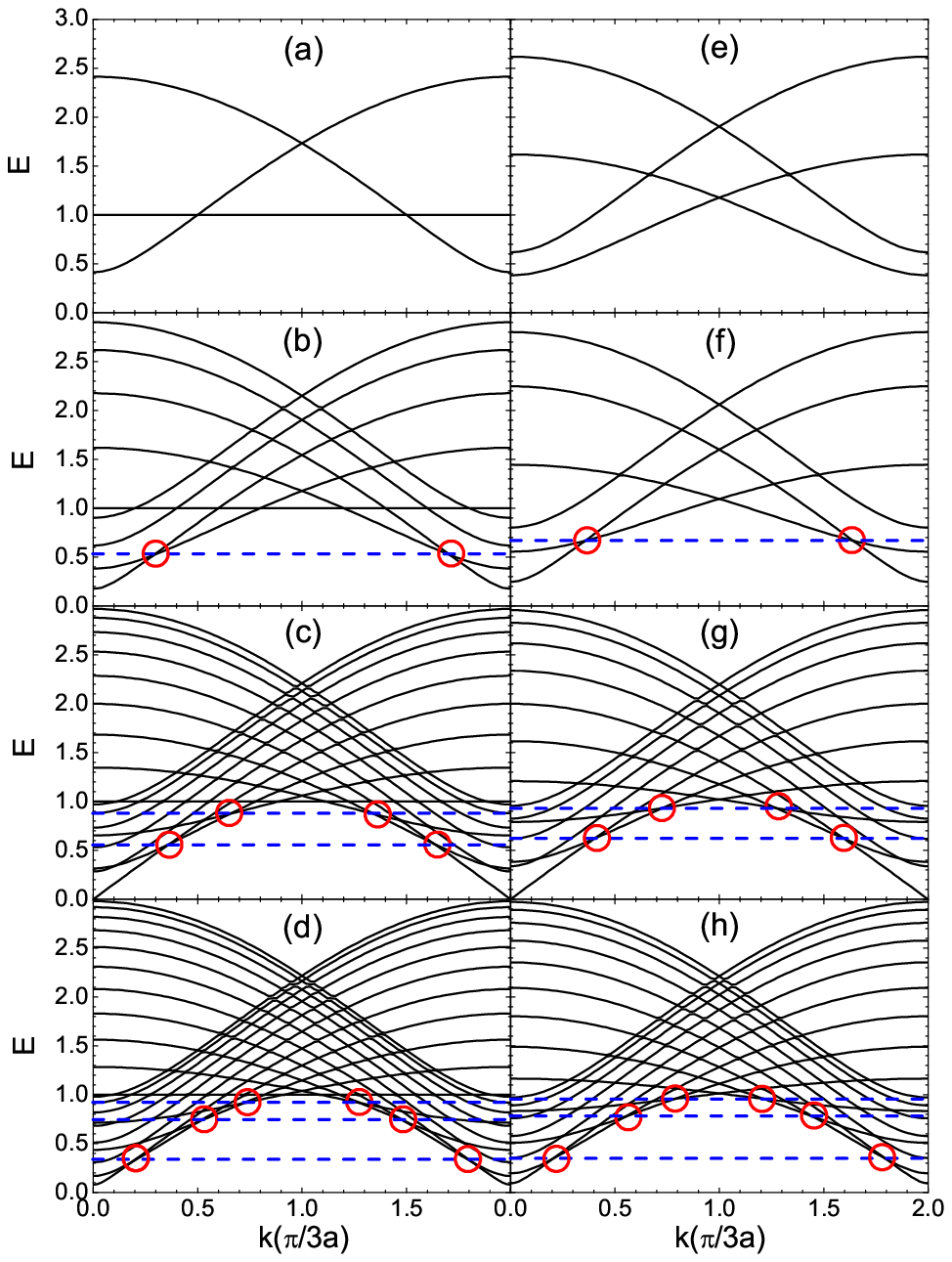}
\end{center}
\caption{Band structure $E(k)$ for AGNRs in widths $N=3(a)$, $9(b)$, $17(c)$,
$21(d)$, $4(e)$, $6(f)$, $14(g)$ and $18(h)$. The magnetic flux is taken as
$f=1.0\times10^{-4}$. The blue dash lines and the red circles indicate those
crossing points which induce the sign reversals in Hall conductivities. }%
\label{Liufig2}%
\end{figure}

However, for AGNRs, besides the band-crossings near of
$E=\pm\left\vert t\right\vert $ there might emerge some extra
band-crossings. The number of crossing points depends on the width
of ribbons. In fact, the AGNRs can be divided into two classes,
odd and even AGNRs, respecting to their widths $N=2n+1$ and
$N=2n$. One of the different characteristics between them is that
there presents a full flat band at $E=\pm\left\vert t\right\vert $
only for odd AGNRs (Fig.\ref{Liufig2}(a)-(d)). Such a flat band is
absent if widths $N=2n$ (Fig.\ref{Liufig2}(e)-(h)) due to
reflection symmetry breaking with respect to the transversal
direction. An immediate consequence of this classification is the
density of states at $E=\pm\left\vert t\right\vert $ to be
different for two-class AGNRs. Hence, the conductivities display
precisely very different when Fermi energy sweeps over
$E=\pm\left\vert t\right\vert $. This characteristics implies that
there is another index in classifying the AGNRs beyond the
distinction between metallic and semiconductor by $N=3m-1$ or not,
where $m$ is an integer. In general, the $k$-independent flat
bands at $E=\pm\left\vert t\right\vert $ become $k$-dependent in
the presence of magnetic field. Therefore, we have to investigate
the Hall conductivities for odd and even AGNRs separately.

\begin{figure}[ptb]
\begin{center}
\includegraphics[bb=15 17 286 316, width=3.205in]{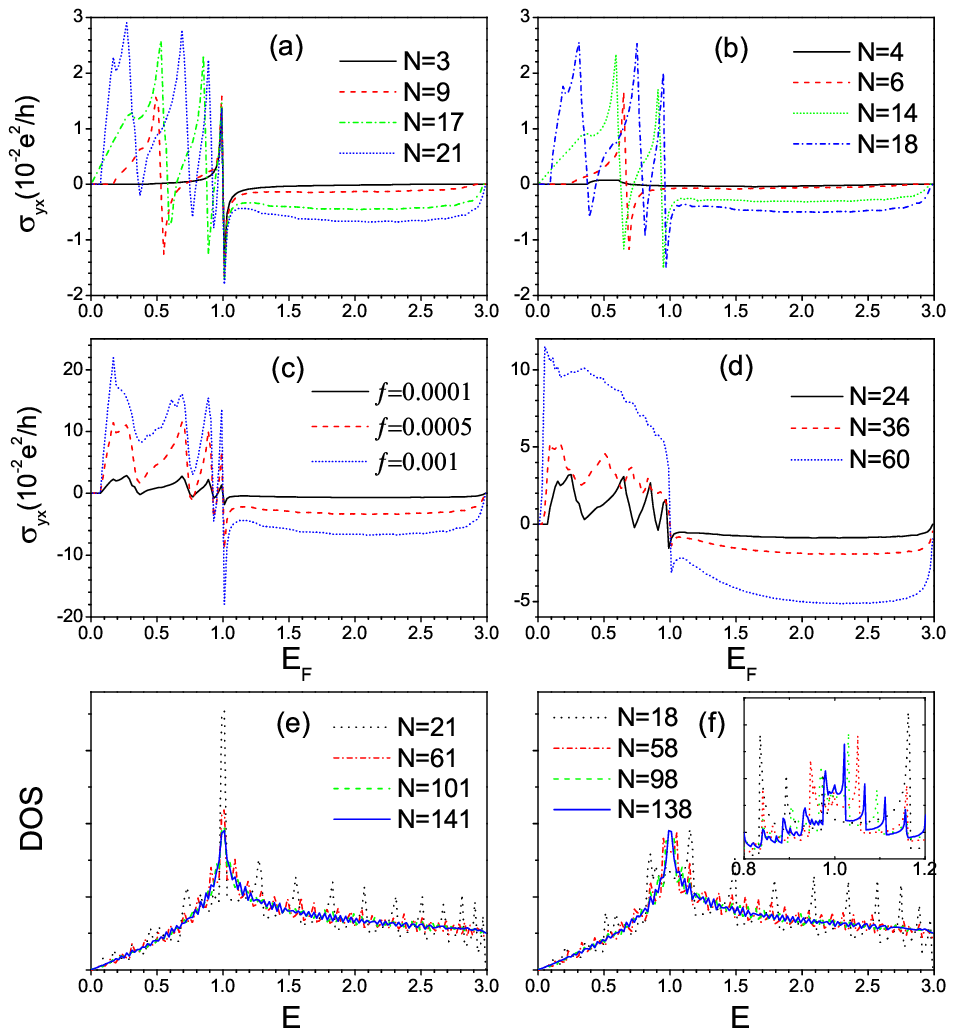}
\end{center}
\caption{The Hall conductivities at zero temperature versus the
Fermi energy for AGNRs in various widths (a) ($N=3$, $9$, $17$,
and $21$) and (b) ($N=4$, $6$, $14$, and $18$), with the magnetic
flux $f=1.0\times10^{-4} $. The evolution of jumps in Hall
conductivities for AGNRs with increasing the strength of magnetic
field (c) ($f=1.0\times10^{-4}$, $5.0\times10^{-4}$, and
$1.0\times10^{-3}$, for a fixed width $N=21$); and with increasing
width (d) ($N=24$, $36$, and $60$, for a fixed magnetic flux
$f=1.0\times10^{-4}$); (e)
and (f) The DOS for odd and even AGNRs in various widths. }%
\label{Liufig3}%
\end{figure}

For $N=2n+1$, it is shown in Fig.\ref{Liufig3}(a) that the flat
band provides a sign-reversal jump in Hall conductivity. In
addition, the energy spectrum shows that there exist $n$ subbands
above $E=\left\vert t\right\vert $ and $n$ subbands below
$E=\left\vert t\right\vert $ at $k=0$ in the region
$0<E<3\left\vert t\right\vert $. The two sets of $n$ subbands
belong to two different valleys. The phase shift between them is
$\pi$. The values of matrix elements
$J_{jj^{\prime}}^{y}J_{j^{\prime}j}^{x}$ among inter-valley bands
are small in the comparison of those among the intra-valley bands.
Thus, the crossing points executed by the two bands belonged to
different valleys contribute to the Hall conductivity weakly
respect to those contribution from the intra-valley. At some
special intra-valley crossing points the values of matrix elements
$J_{jj^{\prime}}^{y}J_{j^{\prime}j}^{x}$ are significantly
enhanced. We evidently affirm these intra-valley crossing points
with dash lines and circles in Figs.\ref{Liufig2}(a)-(d). If we
sort those $n$ bands with energies located above $E=\left\vert
t\right\vert $ at $k=0$ in numbers $j=1,2,3,\cdots,n$, the
crossing points formed between two neighbor subbands
$j=j^{\prime}\pm1$ would lead the sign-reversal jumps in the Hall
conductivity, while contributions from those band-crossings
related two bands $j^{\prime}$ and $j=j^{\prime}\pm m$ with
$m\geq2$ are very small. Fig.\ref{Liufig3}(a) shows that the
sign-reversal jumps in conductivities occur when the Fermi energy
takes the values of energies at special band-crossings. The number
of crossing points caused by two neighbor subbands depends on the
width of ribbons. For those narrow ribbons, the population of
subbands is too sparse to form intra-valley crossings except the
one on the flat band at $E=\left\vert t\right\vert $. With
increasing the width of ribbons, the band-crossing starts to
appear. The number of crossing points is fixed for a certain range
of ribbon widths, so does that of sign-reversal jumps in
conductivities. For example in the range $N=3$ to $N=7$, the
band-crossing occurs only on the flat band. Correspondingly only
one sign-reversal jump appears at $E_{F}=\left\vert t\right\vert
$. If the width is increased to $N=9$, another intra-valley
crossing point occurs at $E=0.52$. A new sign-reversal jump
appears in the Hall-like conductivity. In further, the second new
intra-valley crossing point, satisfying rule of
$j=j^{\prime}\pm1$, appears only when the ribbon is widened to
$N=15$ and the second sign-reversal jump is formed.

For AGNRs in width $N=2n$, there are also $n$ subbands above $E=\left\vert
t\right\vert $ and $n$ subbands below $E=\left\vert t\right\vert $ at $k=0$ as
shown in Fig.\ref{Liufig2}(e)-(h). However, different from the case of
$N=2n+1$, there is no reflection symmetry in the transversal direction and the
flat band at $E=\left\vert t\right\vert $ is absent. The Hall conductivity
would have no sign reversal when the Fermi energy is swept over $E_{F}%
=\left\vert t\right\vert $. Fig.\ref{Liufig2}(e)-(h) show the
band-crossings below $E=\left\vert t\right\vert $ for various
widths. Widening ribbon makes those band-crossings at high
energies to inhabit near of $E=\left\vert t\right\vert $ and be
gathered exactly at $E=\left\vert t\right\vert $ in the limitation
of an infinite graphene sheet. As discussion above, the
band-crossing between two neighbor subbands gives rise to a
drastic change in the Hall-like conductivity. In
Fig.\ref{Liufig3}(b) we have shown the Hall conductivities for
$N=2n$ AGNRs with several $n$. The jumps are attributed to
band-crossings indicated by dash lines and circles in Fig.\ref{Liufig2}%
(e)-(h). It has been seen that, for the narrowest even AGNRs,
$N=4$, there is no intra-valley crossing point in the band
spectrum and no any jump in the Hall-like conductivity, either.
The conductivity is very small. When the ribbon is widened to
$N=6$, the first intra-valley crossing point is shaped and a sign
reversal jump appears in the Hall-like conductivity. Widening the
ribbons to $N=12$ and $18$, the two and three band-crossings
caused by neighbor subbands are shaped below $E=\left\vert
t\right\vert $, respectively. Correspondingly, two and three
sign-reversal jumps in the Hall conductivity are bound to be
followed by the number of band-crossings. It is noticeable that,
along with widening ribbons, the last sign reversal jump tends
toward to $E=\left\vert t\right\vert $.

It is worthwhile to point out that, although the ribbons in widths $N=3m-1$
(integer $m$) are metallic, the number of band-crossings does not change so
long as the width is fixed in some regions. For example, between $N=9$ and
$15$, AGNRs in width $N=11$ becomes metallic, there is still only one
band-crossing point below $E=\left\vert t\right\vert $. The only difference
from those semiconductor ribbons ($N=9$ and $13$) is that the Hall
conductivity is nonzero for metallic ribbons at $E_{F}=0$ ($N=17$ and $14$
have been shown in Figs.\ref{Liufig3}(a) and (b)).

Although the Hall conductivities for ZGNRs and AGNRs of finite
widths have been shown very different, they manifest themselves
the same peculiar odd-integer Hall plateaus as the widths are
widened to transversely infinite. We have seen in
Figs.\ref{Liufig3}(a) and (b) that increasing the width lifts
those sign-reversal jumps in the low energy region. For the
ribbons widened enough, the conductivities become positive for
$E_{F}<|t|$ while negative for $E_{F}>|t|$. The only jump occurs
at $E_{F}=|t|$. As a demonstration, we show shrinking oscillations
with increasing magnetic field in Fig.\ref{Liufig3}(c) and with
widening the ribbon in Fig.\ref{Liufig3}(d). It is well known that
both the band structure and density of states (DOS), $D(E)$, of a
material contain important information about transport properties.
Our numerical calculation show that most of the physical features
appearing in the DOS of the infinite system could be recovered if
the width enlarges to $N\simeq140$ (Figs.\ref{Liufig1}(d),
\ref{Liufig3}(e), and \ref{Liufig3}(f)). The whole DOS profile
evidently shows the Dirac fermion and the ordinary fermion to be
separated at $E=\left\vert t\right\vert $. For an infinite
graphene sheet the DOS tends to constant at $\Gamma$ point
($E=3\left\vert t\right\vert $) due to parabolic dispersion of
ordinary fermions, while vanishes linearly around zero energy due
to the relativistic dispersion of Dirac fermions. As shown in
Fig.\ref{Liufig1}(d), although edge states (Fig.\ref{Liufig1}(c))
give rise to a DOS divergence at zero energy for ZGNRs, it is
deduced as the width of ribbon increase so the edge effect is
relieved. In the view of the restore of van Hove singularities for
even AGNRs in the limitation to the graphene sheets,
Fig.\ref{Liufig3}(f) shows that increasing the width to a certain
region, one can clearly see the emergence of important
characteristics of the DOS of the infinite graphene sheet, namely,
to buildup of the van Hove singularities. Therefore, although the
flat band is absent for a narrow even AGNRs, it is found that the
bands would gather at $E=\left\vert t\right\vert $ in infinite
width, the flat band is built at $E=\left\vert t\right\vert $ and
the DOS becomes singularity. However, the changes of the DOS for
$N=2n+1$ (Fig.\ref{Liufig3}(e)) and $N=2n$ (Fig.\ref{Liufig3}(f))
in increasing $N$ are not the complete same. On the one hand, the
global feature of DOS tend to similar for both odd and even AGNRs.
On the other hand, for the region near of $E=\left\vert
t\right\vert $, they show very different behavior when the width
of ribbons increase. Although those peaks in the regions near of
$E=\left\vert t\right\vert $ are smoothed as the width of AGNRs
increased, the amplitude of the main peak around $E=\left\vert
t\right\vert $ deduces for odd AGNRs, while, on the contrary, a
sharp peak at $E=\left\vert t\right\vert $ is developed for even
AGNRs (as shown in the inset of Fig.\ref{Liufig3}(f)). The
emergence of a peak structure for both cases is expected to be
consistent with the diverging DOS at van Hove singularity for the
graphene sheet. Experimentally, it is expected to observe the
sign-reversal jumps. For ribbons in certain widths, it is hopeful
to build band-crossing at the Fermi energy not far from Dirac
point. For example, for the ribbon in width $\sim2.46nm$ ($N=21$),
there is one band-crossing at the energy $0.8672eV$ and,
correspondingly, one jump appears. Therefore, the quantum
confinement effect on electronic properties for narrow GNRs can be
detected through measuring Hall-like conductivity. The above
calculations are carried at zero-temperature, however, the
fundamental characteristics in conductivities for ribbons does not
changed at the finite temperature. It is found that the amplitudes
of jumps are weakly reduced at the finite temperature.

In summary we have investigated the quantum confinement effect on
the Hall conductivity for GNRs under a magnetic field. It is found
that edge characteristics and quantum confinement of ribbons
affect the electronic transport significantly. Different from sign
reversal jump occurred only at $E_{F}=\left\vert t\right\vert $ in
the Hall-like conductivities for ZGNRs, there are extra drastic
jumps for AGNRs. The occurrence of jumps in conductivities is
attributed to the band-crossings in the band spectrum. The
sign-reversal oscillations would be relieved partly when the width
of ribbons or the strength of magnetic field is increased. The
restore of Hall plateaus and van Hove singularity in the band
structure has been discussed by analyzing the evolution of DOS
with widening GNRs to the graphene sheets.

\begin{acknowledgments}
Authors would like to thank Z.B. Li, B. Rosensterin, and C. Zhang for valuable
discussions. This work is supported by NNSFC and NBRP-China.
\end{acknowledgments}

\end{document}